\documentclass[runningheads]{llncs}
\usepackage{graphicx}
\begin{document}
\title{The Role of Functional Programming in 
Management and Orchestration of Virtualized
Network Resources\thanks{Supported by ERASMUS+ project ``Focusing Education on Composability, Comprehensibility and Correctness of Working Software'', no. 2017-1-SK01-KA203-035402 and the research project ``Reliability and Safety in Complex Software Systems: From Empirical Principles towards Theoretical Models in View of Industrial Applications (RELYSOFT)'' no. IP-2019-04-4216 funded by the Croatian Science Foundation.}}
\subtitle{Part I. System structure for Complex Systems and Design Principles}
\titlerunning{Management and Orchestration of Virtualized Network Resources}
\author{Tihana Galinac Grbac\inst{1}\orcidID{0000-0002-4351-4082} 
%\and
%Nikola Domazet\inst{2}
%
\authorrunning{T. Galinac Grbac }
%\and N. Domazet
\institute{University Juraj Dobrila of Pula,
Zagreba\v{c}ka 30, HR-52100 Pula, Croatia}
\email{tihana.galinac@unipu.hr}}
%\and
%Ericsson Nikola Tesla, Krapinska 45, HR-10000 Zagreb, Croatia
%\email{nikola.domazet@ericsson.com}
%
\maketitle              % typeset the header of the contribution
\begin{abstract}
This is part I of the follow-up lecture notes of the lectures given by the authors at the  
\emph{Three ``CO'' (Composability, Comprehensibility, Correctness)} Winter School held in Ko\v{s}ice, Slovakia, in January 2018, and Summer School held in Budapest, Hungary, in June 2019. In this part we explain the role of functional programming paradigm in the management of complex software systems, and how the functional programming concepts play important role in the designing such systems. Key prerequisite for implementing functional programming concepts is properly designed system structure    following well defined design principles and rules. That is the main goal of this lecture to introduce students with proper system modeling. Furthermore, we also explain how new emerging technologies are designed in such a way that they enforce the development of systems that comply to the design rules inspired by the functional programming. This is extremely important in view of the current network evolution and virtualization concepts, which will require many functional programming concepts in the network services and functions, as will be discussed in part II of these lecture notes.   

These notes provide an introduction to the subject, with the goal
of explaining the problems and the principles,
methods and techniques used for their solution.   
The worked examples and exercises serve students
as the teaching material, from which they can
learn how to use design principles to model effective system structures. Here we focus on students understanding of importance of effective system structures for coordination of development and management processes that are driven by business goals and further evolution.

% ---------------------------------
% iz prezentacije:
% Programming of Management and Orchestration of Virtualized Network Resources. Network Functions Virtualization is a new paradigm for changing the way networks are built and operated. Decoupling software implementation from network resources through a virtualization layer introduces a need for developing sets of NFV management and orchestration (MANO) functions. We focus on coordinating management functions implemented within different functional blocks to accomplish reliable operation for MANO functions operating in distributed environments. The challenges are illustrated on a practical example on Open Stack virtual technology and on the problems inspired by telecommunication industry.

% Focus is on Reliable operation of Management and Orchestration functions of Virtualized resources

%This is the extended abstract of the lecture notes delivered 
%by the first named author at
%\textit{The three ``CO'' (Composability, Comprehensibility, Correctness) 
%Winter School (3COWS)} held in Ko\v{s}ice, Slovakia, on
%January 22-26, 2018.

%
%
%

\end{abstract}

\section{Introduction}
\label{sect:Intro}

%def complex system, functionality, function, req. availability, reliability, safety?

During the last decade of working with undergraduate and graduate students of computing at the Faculty of Engineering, my main reflection on their project work is the lack of understanding what a complex software system is, what and where the problems with complex software systems are, why we need and how to define and design effective software system structures. Students lack the basic understanding of system modelling concepts in general. 

However, on the other hand, from my decade of industrial experience within Ericsson R\&D, I am aware of the importance of these particular skills for software engineers since majority of software industry is increasingly facing with complexity of software systems. The main problem is that students are usually not exposed to large complex systems design and development, and usual student projects are smaller software products built in teams of maximum 1-5 students, in duration of at most 4 months during the semester. These lecture notes are intended for the final year of master study programs or PhD study programs with the main aim to introduce students into better understanding of complexity of modelling and building complex software systems. Therefore, in this lecture I am reflecting theoretical definitions not only to examples of complex software systems but also examples of other complex systems from everyday life, so students may easier grasp the main point of this lecture. Students are asked to think about example system and to apply design principles on these examples. By doing this we engage students to complexity thinking and to reason about causal complexity by promoting discussions around main learnings. Similar teaching approach has been used in \cite{teachingcomplx}. Students are provided with task before each part of course. Student solutions are discussed at the end of each part. Throughout the practical cases we ask students to construct their own view from the content learned applied on the concrete cases of complex software systems but also complex systems in other contexts with reflections on main learning's.

The setting of these lectures is within the theory of complex systems, in particular, the complex
software systems and telecommunication networks. Hence, the lectures start with a gentle introduction to
the theory, carefully positioning the considered problems and challenges within the current evolution
of software systems and networks. Throughout of the lecture numerous examples are provided and discussed with students. Moreover, students are asked at the beginning of the lecture to take a paper and pencil and perform exercises along the course. The course is divided into following main parts:

\begin{itemize}
    \item Introduction to complex software systems, definition of complex system and challenging aspects of their management.
    \item System design principles.
    \item Technologies enforcing design principles.
    \item Reflections on practical examples.
\end{itemize}

Firstly, we introduce the term complex system and relate to complex systems theory. Complex system is defined as system composed of large set of components that interact in between to accomplish specified system functionality. However, global properties of the system that are identified during system execution of previously mentioned functionalities can not be deducted or predicted from the local properties of system components. Exactly, this is behaviour observed in mission critical and large scale software systems that concurrently serves to numerous users. Here we introduce telecommunication network and Mobile switching centre as example of complex software systems and discuss its properties in relation to standard complex system definition. Students are asked to think about example of complex system from everyday life.

Then we discuss main challenges evolving these complex systems. These challenges are mostly related to evolving these systems and delivering its further versions while keeping its global properties in control. The biggest problems arise when we can not predict consequences of introducing changes within the system, and when we can not predict its reaction on environmental conditions. Management of these system becomes extremely costly, requiring lot of expertise, slowly responsive and loosing its competitive business power. In the same time as system grows with number of functionalities, it is loosing on efficiency, number of interactions within the system is exponentially growing, faults are harder to locate, and errors are easier to made. System global properties such are reliability and availability become seriously affected. New paradigms are also needed to accommodate business needs and provide its users grater flexibility in use of resources and on demand. New ideas to increase system scalability are needed. We continue to discuss challenges of system management by using the same examples already introduced in the beginning of this lecture. 

Main challenging obstacle here is human inability to cope with such complexity. Main tool that is used to reason such as systems is \textbf{system structure}. Not only that development projects are using system structure to define work packages, timetables, deliverable and all project activities. But also, human organisation within such software development company is often reflecting the software system structure that is being developed. Consequently, human interactions are usually also reused from system procedures and system behaviour. Thus, the quality of system structure may be a critical for business success. Hence, we introduce main design principles that are used to successfully define system structure and have been already identified to be crucial to cope with system complexity in many other fields. These principles enables easier system management. Here, we introduce students with main system design principles such are:
\begin{itemize}
    \item modularity,
    \item abstraction,
    \item layering, and
    \item hierarchy.
\end{itemize}

Modular are systems that can be easily divided into number of components. In a case of complex software systems the functional system decomposition is followed. System functions are then abstracted and as such provided as a service to its users. Note that system operation is now organised in a service requesting and service providing fashion. Set of functions provided within the system is organised into set of layers. Similar functions are grouped together within one layer that may be developed and replaced independently of other system layers. Also, additional communication rules among these functions are restricting number of possible interactions by introducing hierarchy among system layers. Here we exercise with students their view on possible reflections of these principles on already discussed examples of complex systems.

Although, the main four design principles are well known still there is need for technologies that would directly or indirectly enforce its use. We introduce main technologies that were developed to enforce correct implementation of aforementioned design principles such as Client--server architecture, Service orientation and virtualisation, that are currently widely used within complex software systems and networks. Here, we explain the new challenges arising while implementing these technologies and show students how to deal with them using the available programming techniques.

Finally, in the last section we open discussion on new challenges arising with introduction of 5G (five generation) network. Throughout the course we are examining all the theory on example of telecommunication network but focusing on complex system. At this point students would be ready enough to understand main ideas driving network evolution and how we solved challenges of complex system. Here we will introduce at a high level novelties and challenges of 5G network release and discuss their vision and ideas how to approach them. In the same time this final discussion will be used as introduction to the next lecture on design principles in network evolution and the role of functional programming at network level.

These notes provide an introduction to the subject, with the goal of explaining the problems and the principles,
methods and techniques used for their solution.   
The worked examples and exercises serve students
as the teaching material, from which they can
learn how to use functional programming to
effectively and efficiently coordinate 
management and design of complex systems.

The methods and techniques explained in these 
lecture notes, are already existing and we
claim no originality in that sense. The purpose
of these notes is to serve as a teaching 
material for these methods. Our contribution here is discussion of this topic on telecommunication network example by using industrial experience in developing these systems and previous lecturing experience teaching on this topic to master lever students of computing and during preparatory course provided to the newcomers in industry.

% Your introduction goes here! Some examples of commonly used commands and features are listed below, to help you get started. If you have a question, please use the help menu (``?'') on the top bar to search for help or ask us a question.

% -Razmisli o learning outcomes – sto trebaju znati

% -Koje aktivnosti ce tome doprinijeti
% -Koje alate trebas za to osigurati

% Definicije: Virtual resources, management and orchestration,
% complex systems - evolution and concepts, principles(layering)
% Zasto je vazno to programirati?

% To su najosjetljivije operacije, zato je vazno imati programing konstrukcije

% Primjeri benefita u software engineering-u?
% Diskutiraj slucajeve.

% Programiranje Heat skripti.

% Pitati za ideje gdje bi se ti koncepti mogli pokazati?
% Zamisli kao da s njima svima pises clanak.

% Pitanja
% --------
% 1. What should students know?
% 2. What should student would be able to do with this knowledge? – da bis bila u stanju raditi assessment.

% Outcomes: Remember, Understand, Apply, Analysis, Synthesis, Evaluate

% Sazetak (iz prezentacije): Why and where we need virtualisation? Management functions in virtual environments, The importance of Reliability, Examples 
% Summary

\section{Complex Software Systems}
\label{sect:ComplexSyst}

In this section we will introduce with basic terms and concepts used along within this lecture. Firstly we define complex systems from theory of complex systems and complex software systems. Here we introduce our examples for discussion. We also ask students to provide their ideas for complex software system. Furthermore we discuss challenges arising while developing such complex system. We also ask students to discuss challenges on their own examples as well to provide their viewpoint on the potential challenges. 

\subsection{Systems become more and more complex}
\label{subsect:increased_complexity}

Software systems are built with aim to accomplish some particular end user need. In the last few decades the number of functions the software replaced the human being is continuously growing. Furthermore, the software systems support humans in more and more complex tasks. As result, the software systems are becoming more and more complex and new concepts are needed to cope with their management in order to keep satisfied its users. It is worth to note that these systems are evolutionary developed usually following product line concept like is the case for example in car industry. For example Volkswagen model Golf 1, 2, 3.,...  evolve in sequence of releases and each version has improved engine version but also involve number of additional features.

These complex systems usually involve number of levels of abstraction. System is modeled as number of interacting system components, each specialised for some function.  System requirements are defined at \textbf{global system level} and involves definition of expected system functioning. However, these global system requirements (functional and non--functional) requires intervention and implementation on \textbf{low system level}. So, for every new requirement the global system functioning has to be decomposed into low level system design in order to identify and implement low level details and their interactions needed for functioning of new requirement. As systems complexity grows, it is harder to keep details in global system view that is crucial to understand global system behaviour and its interaction with low level design details. Moreover, it is very hard to understand how the changes in low level details system design may affect global system properties. This is exactly the main characteristic defining \textbf{complex system}. There are number of definitions of complex systems. In \cite{barabasi:Network-science} the complex system is the system with number of levels of abstraction and where there is no clear links between local and global system properties.  

Usually during system design there are numerous possibilities how to design system at low level in order to accomplish global level requirement. Good engineering approach is to identify candidate solutions and based on argumented reasoning to select the best design solution among number of possible solutions. However, number of possible candidate solutions grows exponentially with increase in system complexity. In complex systems management we lack tools and methods for their mathematical modeling and only possible solution is through simulation of their behavior. Systems may be nested within other systems. Then their function may play critical role in functioning systems of systems. Management of such complex systems becomes challenging task.

\subsection{Quality of complex software systems}
\label{sect:Reliability}

Complex systems did not evolve accidentally. Huge effort is invested to develop these systems and lot of programming and functional expertise was needed. There must be a great interest into system functionality, mostly serving numerous users, that lead to further system evolution and that makes these systems to grow into complex system. These systems were usually developed in sequence of projects over decades, by hundreds or thousands of developers and technical experts. These systems mostly perform tasks that are 
of crucial importance for community (examples are in telecommunication, defense, health) for very large number of end users (everybody use them directly or indirectly). In such systems reliability and safety becomes of crucial importance.

Reliability is defined according to IEEE standard glossary \cite{IEEE1990} as ability of system or component to perform its required functions under stated conditions for a specified period of time. This particular requirement has no defined specific implementation implementation reflection. However, any complex system has to implement number of technologies, follow numerous strict rules and procedures that are technology dependent to successfully deliver this particular requirement. Also, numerous verification activities during system lifecycle, that are very costly, are devoted to fulfilment of this particular requirement. This requirement is closely related to ability of the system that is available within specified period of time. System may be unavailable because of numerous implementation limitations such are system failures and congestion. These limitations in most cases arise due to unintentional human mistakes and human inability to cope with system complexity and communication issues when numerous developers worldwide implement the system parts that should work commonly in harmony to deliver specified functions and system functionalities to end users. When system become complex it is hard to oversee all the possible implementation implications on system functioning. Things get very complicated when number of processes that may be active simultaneously for example in threading system use same resources such is corrupted data, files, on irregular interface.

Here, in this lecture, we will focus on well known design principles and concepts that are related to proper system structuring and simplifying its reasoning and management in order to minimise probability of introducing system malfunctioning.

\subsection{Software system structure}
\label{subsect:structure}
When a complex system is constructed, there are numerous possibilities how to design and implement such system. The way how system is built is limiting or enabling its further evolution and system maintenance. For building large scale complex systems, which provide complex functionalities, functional system composition is one logical solution. 

For such complex systems the communication tool among all these parties involved is of crucial importance. \textbf{System structure} is the main instrument to engineer these systems and to connect between this global and local system view. Its main purpose is to enable humans to reason and manage the implementation of the functionality the system will provide to its users. Also, the system structure is very important communication tool among all parties involved within software lifecycle. System structure may also have influence on product documentation and may limit product selling process and companies business opportunities. 

One system is implementing a number of \textbf{system functionalities} for its users. System in operation is accomplishing system functionality by interacting number of system functions. Complex systems usually follow functional system decomposition. Efficient systems have defined a structure of \textbf{system functions} that may serve variety of system functionalities. Therefore, keeping all possible side effects of function change on variety of system functionalities is getting more and more complex and expensive. There, the importance of functional programming paradigm is becoming extremely important as system complexity grows. This means that we tend to treat program execution while operating system functionality as evaluation of mathematical functions without influencing with global system state and keeping mutable data across system functions. 

Principles of writing clean code and function design are to write short functions, well named and nicely organized \cite{martin:clean-code}. In this lecture our focus of interest is exactly the system structure that is logical scheme of connecting global system property (system functionality) to low level system design (system function). We focus on design principles engineering complex system structures. Note, that these principles become necessity when facing with system complexity. Here, we want to describe importance of well designed system structure for system management, further development and reaching business targets. Also, we want to describe how to develop effective system structures capable to deal succesfully with growing system complexity and challenging business needs. 

The system structure provides a blueprint of system modules (usually implementing specific functions), their relations and properties of modules and relations. It is a term related to descriptive documentation of real implemented system. Proper system structure uses hierarchy to logically represent and manage module roles and roles of their relationships. Hierarchy decompose system modules and relationships into several distinct system layers where each layer has its own role in system functioning. 

Currently we lack systematic approach to engineer complex system structures. However, the software systems engineering theory has developed design principles that are used to guide software engineers while developing such a systems. Furthermore, there are architectural technologies developed that promote use of these system design principles. In contrast to system structure, a system architecture is a metaphor similar as architecting a building \cite{architecture} and its main purpose is to provide system blueprint but for the developing project purposes, where we need to derive overview of project tasks necessary to implement new requirements. Further, in this lecture we will address the main design principles and reflect on selected technologies that support their implementation. Our intention is to discuss relevance of these technologies from the perspective of system design principles.  These understandings will be of crucial importance for Part II of these lecture where we will extend the understanding of functional system structuring concepts to a network modeling, management and orchestration.

\subsection{Software organisations developing complex systems are complex too}
\label{subsect:complexorg}

From software engineering perspective the software organisations responsible for development and maintainance of these complex systems are usually very complex, globally distributed and involve number of developers into software development lifecycle. From business perspective, there are number of interested parties into these system and software development usually involves number of stakeholders in requirements elicitation and system definition phase and sometimes with contradicting requests. The key element for developing such a complex systems is concise and structured communication among all involved parties that lead to simple solutions. Software engineering tools, methods, practices and principles are of crucial importance to support complex system development processes.

For example, when an functionality of a system in operation experience failure there are serious consequences for the owner of the system and its numerous users. 
On the other hand side, when the system functionality experience failure it may be hard to identify fault location within the system implementation. The link between global system property malfunction and low level system design has to be identified. However, this link is not always clear and may involve conflicts in interaction among components at low level design. Furthermore, if we consider that components may be developed by different organisational units that are usually globally distributed then speed of finding a solution to the experienced system failure may depend on communication tools global organisations use. The cost of system 'out of order' in this case scenario is seriously impacted with these tools.

Complex software systems have usually numerous markets and applications. Therefore, it may become challenging task to maintain all the systems versions that are running in different markets or in different applications. For example, when system has many variants, each tailored to national specific regulations managing all configuration versions may become a challenging task in the case when system is not modular. Furthermore, when failure occurs a fault mapping process becomes extremely expensive for all system variants.

Organisation that own monolithic in house developed complex systems have to reconsider their business goals. There may be considerable specialist knowledge invested into their development that is proprietary right of the owning organisation. On the other hand side, business drives, like cost of ownership, cost of change may force organisation to open part of its product that may interact with system functions developed by other organisations by using open standards. Thus, organzations are forced to introduce competitive approach not only at system boundaries but also within systems internal structure.

\subsection{Software engineering is not mature discipline}
\label{subsect:swengdiscipline}

Current knowledge in engineering complex systems, about concepts and theories how to design, maintain these systems is still not mature enough to successfully cope with such complexity. That is why software engineering science behind engineering complex systems is very active and trying to develop new theories that can explain fundamental system behaviour, find adequate models for interconnection of global and local system properties, and trying to find new concepts for better complex systems management. We need this understanding to better reason these systems and thus enable their further evolution. Moreover, fundamental is to understand system behaviour over time and not just observing individual events. With this knowledge we may be able to better engineer autonomous and self management system principles which may turn complex systems into equilibrium stage directed towards business goals. Another aspect of system engineering is to understand system change in respect to introduced time delays. Introduced time delays may completely seriously damage healthy functioning of the system \cite{Thinking-in-systems}.

Let us compare available knowledge in mechanical engineering in engineering vehicles and available knowledge in engineering software systems. We can model and predict changes on vehicle functioning while we are changing its local properties like for example piston dimension. On the other hand side we can not model and predict complex software system behaviour, i.e. if we change a single line of code within complex software system, we can not predict the consequences it may introduce in system operation.

\subsection{Exercises}
\textit{\textbf{Exercises for students to asses learning's from the chapter just read }
\begin{enumerate}
    \item Define the term complex system.
    \item What are essential characteristics of complex software systems?
    \item Define reliability aspect of software systems and discuss reliability requirements in relation to growing system complexity. 
    \item Define software structure.
    \item What are the challenges development organisation is facing when developing complex software systems?
    \item What we mean when we say that engineering complex software system is not mature discipline?
\end{enumerate}}
\textit{\textbf{Exercises for students that requires students to reflect on major issues and to asses their understanding of complex system modelling challenges}
\begin{enumerate}
    \item Lets firstly imagine an example of complex system from our environment. It is not mandatory to be an existing computer system of software implementation. It can be anything that you can imagine, and can fit into complex system definition. Please elaborate in sense of complex system definition how your example can be categorised as complex system.
    \item Then can you think about its possible users and functionalities that system perform. Please write at least three different kinds of its users and at least five system functionalities.
    \item If we have to engineer such a system can you imagine components/functions needed to perform system. Please depict high level system structure, with which you can explain functioning of previously mentioned five functionalities.
    \item List main non--functional requirements your system may have.
    \item Can you imagine how big this system might be? Can you imagine how much developers are needed? Can you think about structure of human organisation that is able to develop such product?
    \item Describe challenges the system might face when introducing new kind of users, new functionalities, massive traffic.
\end{enumerate}}

Note that these questions are used in the classroom to engage students into complex systems thinking and reasoning. Students are asked to answer on these questions on the paper. Then, teacher collects papers and starts group discussion for each question. From my experience the biggest challenge for students is to think about system structure, organisation and the challenges an organisation may face evolving these systems. So, conclusion to the discussion is followed by providing the examples of complex systems that is provided in the following section.  

\section{Examples of complex systems}
\label{subsect:examples}
 
In this section we will provide three examples of systems that we may consider as complex. 

Firstly, we provide an example of human body that is complex system inspired from the nature. This system is not built by humans but the body structure is used when we want to understand its behaviour with main motivation for medical purposes. In that sense we may consider medical science as engineering of human body. Here, we want just to represent how much complex one system may become and that number of functionalities and level of its autonomicity in number of contexts extremely increases their complexity. Since, here we focus on complex software systems we will continue discussion on programming human body and switch to the example of humanoid robot.

Another example we provide is vehicle. Vehicles are mechanical systems built by the humans so high level of engineering was already applied during its development. In contrast to human body, all behavioural processes are modeled by solid mathematical and physical models that can be used to deduct from global to local system behaviour. Engineering of these systems is not considered so complex as in the case of human body. In medical science we still lack adequate models for modeling this global--local interactions to understand human body behaviour. 

Third example we provide is one of the largest and most complex technical system that human has engineered -- the global telecommunication network. Majority of people across the globe use its functionalities in some way to interconnect people. Furthermore, every day we are witnesses of development of new technologies. All these new technologies finds its applications within the technical systems that are all interconnected via telecommunication network. Therefore, telecommunication network becomes one of the key enablers for development of societies. Its great importance lead to its rapid development and rapidly increasing complexity. Its main constructive ingredient is software and as software complexity has grown new innovations were needed for structural approach. Here, number of telecommunication principles were introduced to organize and structure telecommunication functions within the network, and within the telecommunication software. This is why this example is the main leading example we provide across this lecture.

In a sequel we will try to discuss aforementioned exercises by having in mind these three examples of complex systems. 

\subsection{Example 1. Human body and humanoid robot.}

\begin{figure}
\centering
\includegraphics[scale=0.45]{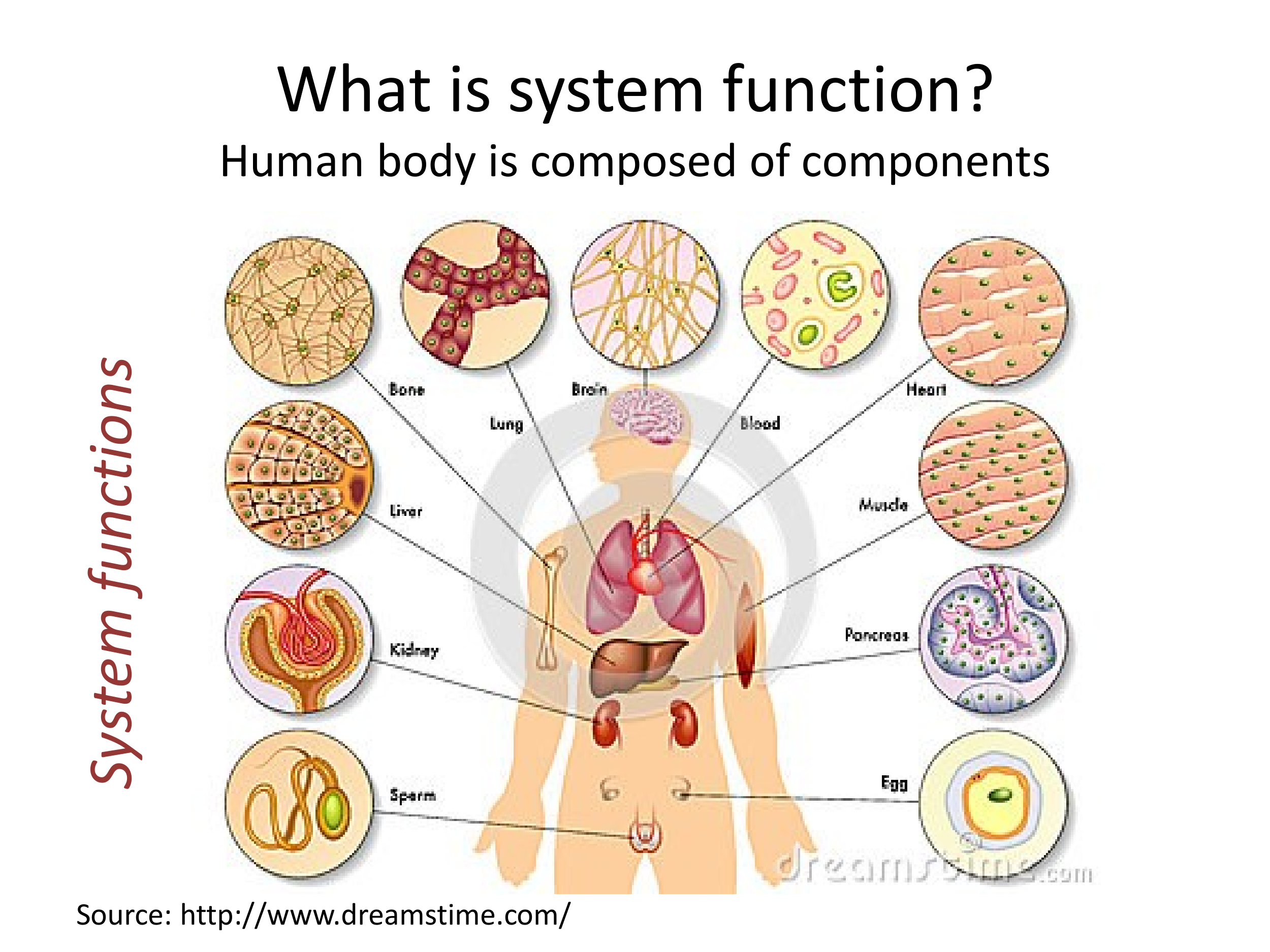}
\caption{Human body, modular structure of system functions.} \label{Humanbody}
\end{figure}

Lets take a human body as analogy for complex system. Humans in their every day life perform number of functionalities; they speak, sing, walk, run, write, perform number of complex tasks by combining its basic functionalities. This human outside view is presented to its environment. On the other hand, from the human inside view the human body is performing a number of body functions in which the body organs take vital role. Human body has very clear structure that is composed of organs and each organ have its unique role in performing human body function. Thus for example the organs within the human body are heart, liver, brain, skin, see Fig. \ref{Humanbody}. Within the human body we have system structure at different levels of abstraction (e.g. cell, etc) and it is very complex to reason among the processes that are executing among different levels of abstraction. Furthermore, there exist communication channels among human body organs that are mandatory in body functioning. The human body communication system may be considered nervous system which is responsible to transfer information across the human body, communicate central brain system with peripheral elements, coordinate actions and sensory information.  

In sense of complex system definition, there is huge gap in understanding of human body global functioning and its relation to local body organ functioning. There may be numerous root causes in malfunctioning of body organs that cause effects on human body functionality. For example, inability to speak may be rooted in some specific malfunction in tongue organ, nervous system, brain function, etc. Furthermore, medical treatment may have numerous side effects and we are not able to predict and control them.

In the example of humanoid robot, let us suppose that we have to program software for robot mechanics. Then, lets suppose that our robot will be a soccer player. Its main functionalities would be to walk, to be able to direct a ball in desired direction, to coordinate with other robots and within the field, to be able to recover from unexpected events such is for example walking on non-perfectly flat surface. An robot that is capable to play a soccer game have to implement number of drivers for all its sensory system needed to interact with its environment. Furthermore, it is implemented as distributed control system where various controllers are used to control all the mechanics needed for its movement. This robot in order to fulfill its required functionalities must have number of different functions. Some functions that may be implemented are control functions for kinematics, coordination of sensory functions, navigation, localisation, etc. In this example numerous effects from environment may effect on robot behaviour and the number of possible scenarios may increase so all robot functions management and related coordination and control functions may become complex. This means that it may become very hard to isolate events that cause robot malfunctioning. Solid robot design would also imply modular robot system with automatical detection and configuration of new sensory systems, new functions and new platforms \cite{Robot}.

\subsection{Example 2. Vehicle.}

Another example of modular system that comes from mechanical engineering is vehicle. It is assembled of number of parts each performing its function needed in vehicle functionality. In the context of software system the analogy to human organs or car parts are entities, modules, system units. Although, there exists whole theory how to build functioning vehicle and how we can structure it, in software engineering there is lack of theories how to build complex software systems. Here we may observe the difference with complex system.    

The systems modules communicate to each other in accomplishing particular system functionality. For these communication purposes there are established physical connections between the system organs. To transfer of particular information between system modules that is needed to accomplish particular system functionality these physical links are used. In analogy to human body these physical links are nervous and communication system is nervous system. Or in mechanical sense of car these are number of mechanical or electrical transmitters. Modern complex software systems are build over the network and system modules communicate over the telecommunication network. Therefore, here we have interleaving of network and software engineering theories to build modern complex software systems.  
 
The system structure provides a blueprint of system modules. It uses hierarchy to logically represent and manage their roles. Hierarchy decompose system modules into several distinct system layers where each layer has its own role in system functioning.\newline

\subsection{Example 3. Mobile Switching Centre within telecommunication network architecture.}

Telecommunication network is set of distributed computer systems, involving hardware and its related software, interconnected with links. Network serve to numerous users that connects to network via various terminals that are distributed geographically, with diverse communicating needs, developed by numerous producers. These systems are all developed and engineered by humans. However, the complexity of this system is much higher then is in vehicle example. The network has to integrate variety of terminals developed on different platforms thus act as interconnection among various technologies, industries, equipment producers. That is why clear and open standards has vital role in further network evolution.

Telecomminication network has evolved in sequence of releases. Its evolution is standardized in various telecommunication standards. This was important to enable interwork among equipment of various produces but also to open competition among network equipment suppliers. Initially, it was built for voice traffic, and further extended to carry data, video and multimedia traffic with very different transport needs and offering variety of services. Also, huge network revolution was introduction of mobile users. Because of all that, network becomes too complex and during its evolution number of structural changes have been introduced and that force redesign of network architecture. These structural changes where always introducing design principles that we provide here at some network abstraction layer. All these structural changes were followed by standardisation bodies. Here we will reflect on work within 3rd Generation Partnership Project (3GPP) that covers cellular telecommunications technologies, including radio access, core network and service capabilities, which provide a complete system description for mobile telecommunications. An excellent overview of 5G mobile communications technologies is presented in  \cite{osseiran-monserrat-marsch:5G-mob-wireless-comm-tech}. From 3GPP specifications it can be observed mobile network architecture evolution across releases 2G, 3G, 4G and finally 5G. Also, here we will explain its implementation of concrete examples.

Main functionalities introduced in 3GPP evolution steps are following:
\begin{itemize}
    \item 2G - Mobile core network is introduced for  GSM (Global System for Mobile Communications) users and voice based services offered by GSM network. The main network functions are located in Mobile Switching Centers (MSC), Home Location Registers (HLR) and Visitor Location Registers (VLR). 
    \item 2.5G - Mobile Core Network is extended for GPRS (General Packet Radio Service) users and data based services offered in GPRS network.
    \item 3G - Mobile core network is extended for UMTS (Universal Mobile Telecommunication System) users and intergated voice, data, video and multimedia (combination of aforementioned traffic) traffic. Integration of GSM, GPRS and UMTS services within the core network is achieved in IP Multimedia System (IMS). Details of IMS system architecture and main design principles and technologies used may be found in \cite{book:IMS}.
    \item 4G introduce concept of Long Term Evolution (LTE) mainly concerns on new core network architecture redesigned to enable rapid evolution by introducing common mobility management functions for all its users and packed based transport for all services.
    \item 5G introduce new network management architecture where rapid network softwarisation is forcing service orientation by offering all of network resources as a services to all network users. As stated in \cite{osseiran-monserrat-marsch:5G-mob-wireless-comm-tech} 5G will create the conditions where wireless connectivity will become necessity in a huge number of applications. 
\end{itemize}

Along this evolution mobile core network architecture has been restructured by following design principles that we will present in following section. Here in this lecture we will explain application of design principles during mobile core network evolution. For that purpose we will focus on design of central network function that is switching of mobile subscribes. Here we will use example of complex software system Mobile Switching Centre (MSC) and its Ericsson implementation on AXE platofrm. Note that switching relates to establishment and release of connection across the network among two end users, connected to that network, which want to communicate. Ericsson's MSC product was implemented on Ericsson proprietary product AXE \cite{PLEX}. AXE is 40 years old in-house developed platform. 

From the beginning product was traditionally developed in monolithic fashion. That means that all node functions where developed in-house by Ericsson, on Ericsson AXE proprietary platform \cite{AXE810}. Most node functions are implemented in software and majority of software is written in Ericsson internal Programming Language for Exchanges (PLEX) from which Erlang evolved, an special--purpose language, concurrent and for real time applications. The system structure followed implementation view, where modules performed specified implementation functions. As product matures, the number of functions grows, and high expertise was needed to further develop and maintain that product. Adding new functionality in already mature product become very inefficient and costly.

AXE based MSC has evolved within more then fifteen releases, has several millions lines of code, is developed in globally distributed organisation involving more then ten development units geographically distributed across the globe \cite{UnderstanTele,AXE810}. For each project release there are several hundred even thousands software engineers involved in its development. Product has requirements to handle concurrently more then one million of users with very strict reliability requirements \ref{sect:Reliability}. Product should be able to minimise delays caused due to serving numerous users concurrently. Furthermore, the switching equipment have to have high level of availability that is directly connected to the system architecture and software structure.  Software is structured into functional blocks (functions like callee and caller number analysis, charging, switching, etc.) with clearly defined interfaces. 

\subsection{Exercises}

\textit{\textbf{Exercises for students to asses learning's from the chapter just read }
\begin{enumerate}
    \item What are the main differences among the examples of complex systems provided in this chapter?
    \item Order the examples of complex systems by level of complexity from the less complex to the most complex. Explain your ordering criteria.
\end{enumerate}}
\textit{\textbf{Exercises for students that requires students to reflect on example of complex systems and discuss on complexity implications}
\begin{enumerate}
    \item
Let us consider for example a Mobile Switching Centre node that is serving node for switching mobile calls. This system has possibility to handle one million of users simultaneously. Imagine that this system experience failure such that it requires restart. Or imagine that system is not modular and adding of new functionality requires system restart. All mobile services, calls, SMS would be discarded. If we want to measure cost of loss for the operator owning this system we may multiple cost of call per minute with number of users and with number of seconds system was out of service. Please calculate operator loss for 1 minute out--of--service at 50\% load. Note that MSC system is very complex and system restart may take hours. This example has the main purpose to increase awareness of students on importance of flexible and reliable system operation and how it may become important in complex systems that serve number of users. 
\end{enumerate}}

\section{Design Principles}
\label{sect:Design:Principles}

Whenever a complex logical problem has to be engineered, there are some typical patterns arising in all computing fields, as for example is the case in logic and programming languages, software engineering, networking, database architecture, etc. Specialists from different fields have experienced the same problems facing complexity, and have come to solutions that are grounded on the same concepts, but perhaps implemented differently in various fields. Therefore, we define here these common concepts in form of design principles: modularity, abstraction, layering and hierarchy. These principles are explained in detail in every serious textbook presenting system design principles \cite{vliet:softw-eng,saltzer-kaashoek:princ-comp-sys-design}.

\subsection{Modularity}
\label{subsect:modularity}

The most frequently used approach when dealing with complexity is division of the system into set of smaller components, simpler to understand, that are independent of each other. This concept is very much used in all fields and we call it \textbf{modularity}. In complex software systems we use functional system decomposition when the system is decomposed into set of independent function units called \textbf{system modules} which interact among each other aiming to accomplish system functionalities. Thus, we have relation between global system level on which we have functionalities that system is able to perform, and local system level where we have structure of system modules which are able to perform specific functions. System functionality is achieved as interwork of set of system functions.

The benefit of system modularity does not only lies in better understanding of system functioning by travelling between global and local system view but also its benefits are seen from perspective of easier collaborative design and expanding business opportunities. Modular systems may be developed by globally distributed organisations and the system responsibility may be shared among system modules and its development organisations in different countries. Also, some modules may be easily given to third parties for further evolution. This way, decreased cost of development have to be carefully balanced with impact on system quality. 

\subsection{Abstraction}
\label{subsect:abstraction}

Abstraction is term very much tied with modularity concept already explained above. It is related to concept of introducing communication rules within the system by introducing standard interfaces among system modules and separation of interface from module internal details. Introducing term standard for interface means that everyone using that interface use same rules of operation. Thus, we have two or more independent modules tied with same interface which may evolve independently while they are interconnected in between with standard interfaces. Idea is to abstract function of each module by using its interface. In other words, the interacting module does not have to know implementation details of other component and their interaction is achieved through exchange of standard set of messages and data.

Additional benefits that arise from abstraction on modular system are numerous like easier system evolution, inherent and autonomous failure prevention and failure diagnosis. 

We can divide software system in number of different ways but the best one is the one where we can use abstraction. For example, in object oriented programming some programming languages have implemented concept of inheritance to force programmers to make their programs to be not only modular but abstract too. This means that for example different geometrical figures that may be drown on GUI like are triangle, square, circle are inherited from the same class called Figure that is an excellent candidate for abstract class, \cite{OODesign}. Thus, functions draw and erase may be defined on Figure object type while exact geometrical object (triangle, square, circle) will be called during program execution, when needed. Separation of implementations in function draw, erase and inherited classes triangle, square and circle is achieved through definition of abstract class Figure. This is very important for limiting the impact of propagation of fault and its effects. When a module is exhibiting a fault, it means that module does not meet its abstract interface specifications. Furthermore, here we allowed various objects with different internal implementation to share the same external interface. This is well known concept called polymorphism in object oriented programming.

Design and implementation of standard interfaces is about definition of set of rules for information exchange among two interacting entities. All possible interactions have to be developed in advanced but also all not allowed and irregular interactions have to be considered and proper action have to be defined in a way that lead to regular system behaviour. Module implementation have to take care for unexpected behaviour as well and avoid system failure situation by securing proper system reaction. For example, if an unexpected message or data is received through standard interface, the component that experience such condition have to implement regular process termination while keeping in mind to release all unnecessary resources. In concurrent systems if one process fails we want to keep all the other active processes. When failure occurs, troubleshooting process of identifying fault location in modular system is much easier then in non modular system. System may be divided into components but not compiling to modularity principle of module independence and function abstraction. Then, there is no guide to control low level implementations and fault may be located anywhere in the system. The role of standard interfaces is to force well defined design rules for implementing interfaces within the system. Trace of exchanged messages over these standard interfaces in execution paths that leads to failure helps in fault isolation process and may detect module that did not comply to standard interface rules. Moreover, every module has to correctly implement regular use cases of standard interface but also have defined actions for irregular use case scenarios. Thus standard and well defined interfaces helps in fault prevention process. 

Furthermore, modules which have clear function within the system may be easily changed and updated without affecting each other until strict interface rules are followed. Thus, the same system may be collaboratively developed and responsibility over the system modules may be distributed across the globe.

In evolving systems there may be number of system versions implemented and in operation and used by numerous users. Development impact on one system version which share components with previous system version are made invisible to all previous system versions until strict backward compatibility rules for protocols on interface are followed. Thus, at system evolution, impacts due to new or improved system functionality are easily located and implemented. Only modules that needs intervention are opened and improved. Change is simple until the change is keep within the module without change of interface. Bigger changes which require intervention on the interface have to impact all components which use this interface. However, interface is also subject of versioning and all new impacts have to comply to strict rules to keep backward compatibility on interface with all components that use that interface but are not affected by change, i.e. new functionality is not implemented. In some cases, the change on interface may involve impact on all interacting units but in that case modularity process was not performed correctly. Note that there exist strict rules how to change modules and interfaces in modular system design if we want to benefit from modularity concept.

\subsection{Layering}
\label{subsect:layering}

\begin{figure}
\centering
\includegraphics[scale=0.45]{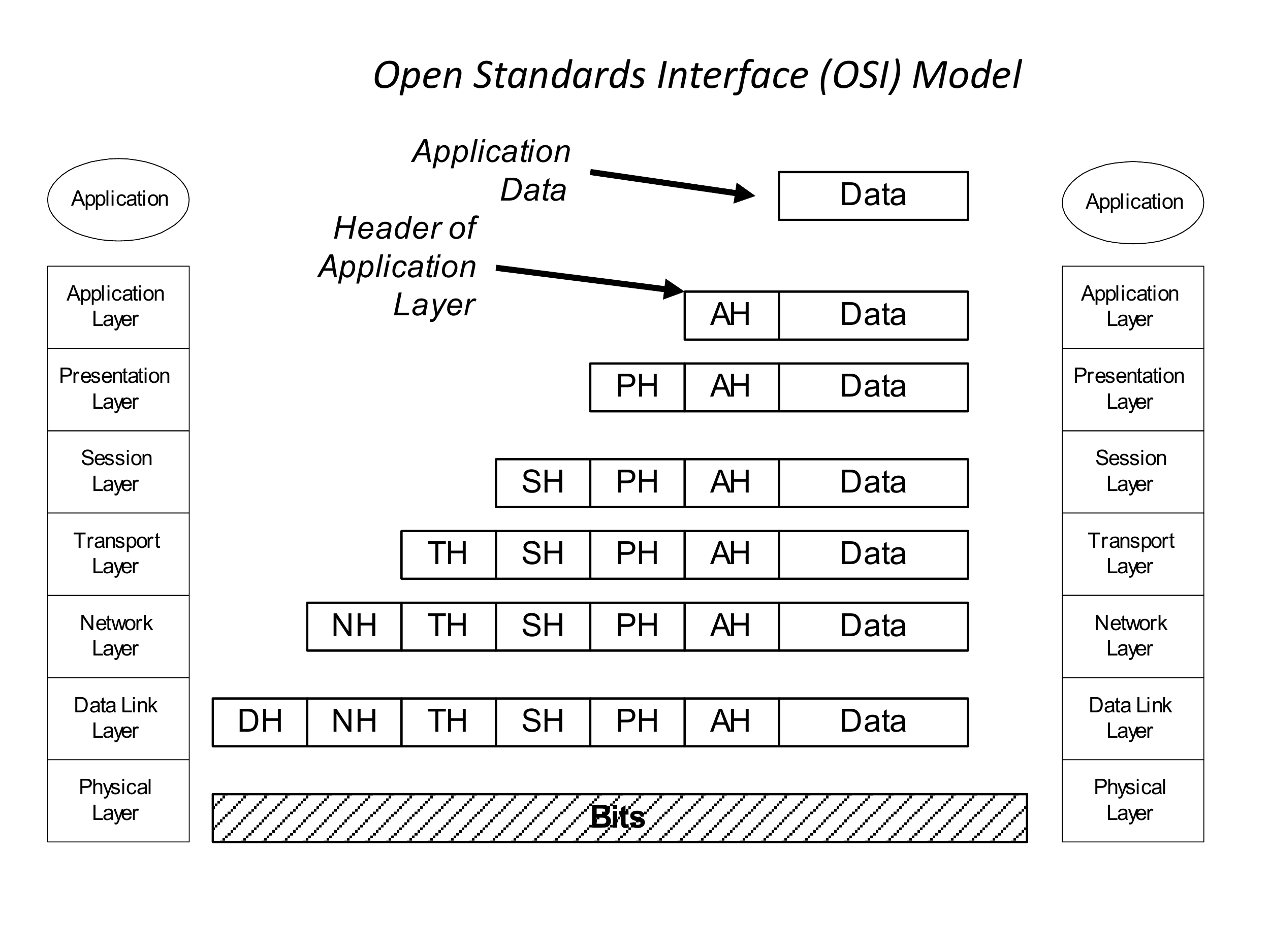}
\caption{Layering communication functions: Open Standards Interface OSI Model.} \label{fig:OSImodel}
\end{figure}

In dealing with complexity we usually use layering. It assumes grouping of functions and related information into distinct layers of abstraction which communicate with standard interfaces by limiting the rules of interaction among various functions. The need for layering in communicating computer systems was recognized early 70ies when various computer manufactures developed their own products that needs to communicate. To avoid situation that each manufacturer developers its own communication rules, the standard body International Organisation for Standardisation developed Open Systems Interconnection Model OSI that made it possible for system produced by different manufactures to communicate in between. One example of similar standardisation effort is international standard IEC 60906-1 that define power plugs and socket-outlets with the main purpose to be used across the national borders. Thanks to these efforts today we do not need specific plug for each country while travelling except for some specific countries. For those countries we need specific adaptors in order to use equipment produced for European market in their countries. OSI Communication model was developed with similar aim. 

OSI Model introduces layering of systems functions in communication with other systems. OSI Model provides details what information is carried by each layer and not describing its implementation details. Thus, the data flow among computers may be followed across the layers of computers involved in communication. The model is depicted in figure \ref{fig:OSImodel}. It consist of seven functional layers. Each functional layer have defined communicating rules with its pear at the same functional layer but within two computers in communication. These layers are physical, data link layer, network, transport, session, presentation layer and application layer. Thus we obtain layering of communication within the network of physical computers and the rules of communication are defined within various standard protocols at each functional layer. These protocols define horizontal communication in computer network. In the figure \ref{fig:OSImodel} it is presented how each layer adds its communication overhead into transmitted data between two computers. Within the computer network there exist communication network at each horizontal functional layer. Furthermore, vertical communication, the communication between the layers remains within one computer and thus may be subject of manufacturer internal standard rules. However, proper design and implementation of functional layers within the system should aim to make the layers independent one of another. Thus, each layer may be changed and reused within the computer architecture without affecting other layers. Such design involve introduction of services within the computer and layer functions may be offered to other layers as services. The service interface should be made clear and independent of other services. Pure functional layering should made layer user unaware of service implementation details. Here, if we are able to introduce standard rules into communication among the functional layers we will be able to use equipment of different producers also at each functional communication layer. Therefore, further network evolution involves splitting of communication architectures and often introducing standard vertical protocols. For example in 4G mobile network architecture a new network architecture was proposed with idea of splitting the network architecture into thre planes applications, control and resource network plane. Gateway Control Protocol, GCP (https://www.itu.int/rec/T-REC-H.248.1-201303-I/en) was introduced in between control and resource layer. The protocol is used to manipulate and control physical resources placed in resources plane needed in mobile end-to-end communications. Another example is introduction of Software Defined Network in 5G network architecture \cite{osseiran-monserrat-marsch:5G-mob-wireless-comm-tech} and OpenFlow logical switch (https://www.opennetworking.org) to control network resources in network plane. Both network layering occurred at different abstraction layers, layering in 4G occurred on virtualisation of network physical resources and layering in 5G occurred on virtualisation of network functions.    

For better understanding of OSI communication model we use postal analogy that is provided in excellent student book \cite{UnderstanTele} where two project managers from different countries communicate to each other on a secret project using the land post infrastructure. OSI functional layers are compared to functional layers involved in communication path when the postal office is used as a service for transferring communication messages.  

\textit{Postal office analogy}
Lets suppose that two project managers work on the same secret project and each manage team in its own office. In a given example in \cite{UnderstanTele} project manager Erik is Swedish working in office in Lule{\aa} and the other Louis is French working in Goteborg. In real telecommunication network, project managers are applications that communicate to each other using telecommunication network which in our analogy is represented by standard land post. 

Since the project is secret project managers agreed to communicate on English, use standard land post for message exchange and since the project is secret they agreed to use encryption of messages. Also, their secretaries exchange addresses and letter formats for the communication. These agreements are equivalent to horizontal layer agreements on protocols to use at each horizontal layer.  These three agreements in our analogy represent agreements used on peer protocols among the functions of same layers in different nodes. Thus, we have project managers and the translators at the seventh layer, crypt--experts at sixth layer and secretaries at layer 5. 

The communication path from project manager Erik in office in Lule{\aa} and the project manager Louis in office in Goteborg is passing all the seven OSI layers once in office in Lule{\aa} and again all the seven OSI layers in office in Goteborg. When project manager Erik writes a letter in Swedish, it gives this letter to translation office to translate it in standard communicating language (in this case English). Then, the letter is transferred to the crypt--expert to encrypt the letter using agreed encryption for the communication with Louis located in Goteborg. When message is encrypted the letter is transferred to the secretary to prepare the letter for transmission through land post. This means, that secretary put the letter into an envelope and address the letter to Louis (its secretary) in Goteborg. When the letter got its addressed envelope it is ready to be passed to the local postal office. Local postal office in our analogy is the entering point to the post network. Here it starts packaging of this letter into postal packets that travel to the same destination. The letter of Louis now is packed with all other letters that have destination to Goteborg postal office. This function is equivalent to the layer 4 functions. At layer 3 and 2 the group of letters with destination to Goteborg postal office is packed into separate packet and mark the address of postal office in Goteborg on that packet. The packet is then hand over to the transporter.  Finally, the transporter is transferring this packet to the Goteborg postal office using pubic transport infrastructure. When the packet is received at the post office in Goteborg the reverse process of passing through layer 1 to 7 starts. Firstly, the packet is opened and the letters are regrouped to the local destinations in Goteborg. Thus the Luis letter in Goteborg postal office is now grouped with all other letters addressed to Luis office that are received from various postal offices. This function is equivalent to layer 2 and 3 at the other communicating party. This group of letters including Louis letter is given to postman that is delivering the letters to the Louis project office. Postmaster employed in Louis project office is delivering the letter to Louis secretary (layer 4). At layer 5 the secretary signs for the received letter, notice that the message is encrypted and transfer the letter to the crypt--expert. Encrypted message the crypt expert passes to the English--French translation (layer 6). Finally, the letter is received to Louis.  

This example we used to better explain layering of functions and communication protocols in computer networks. There, by the standard we have defined seven distinct layers of communication where communication functions are grouped and organized into distinct layers. Furthermore, the communication is allowed only between neighbouring layers and it flows vertically from the layer 7 to layer 1 in one computer then it is transmitted over the physical wires to the other communicating computer where it again flows from layer 1 to seven. Finally, dialog is achieved in between applications located on the top of seven OSI layers in two communicating parties/computers.  

\begin{figure}
\centering
\includegraphics[scale=0.45]{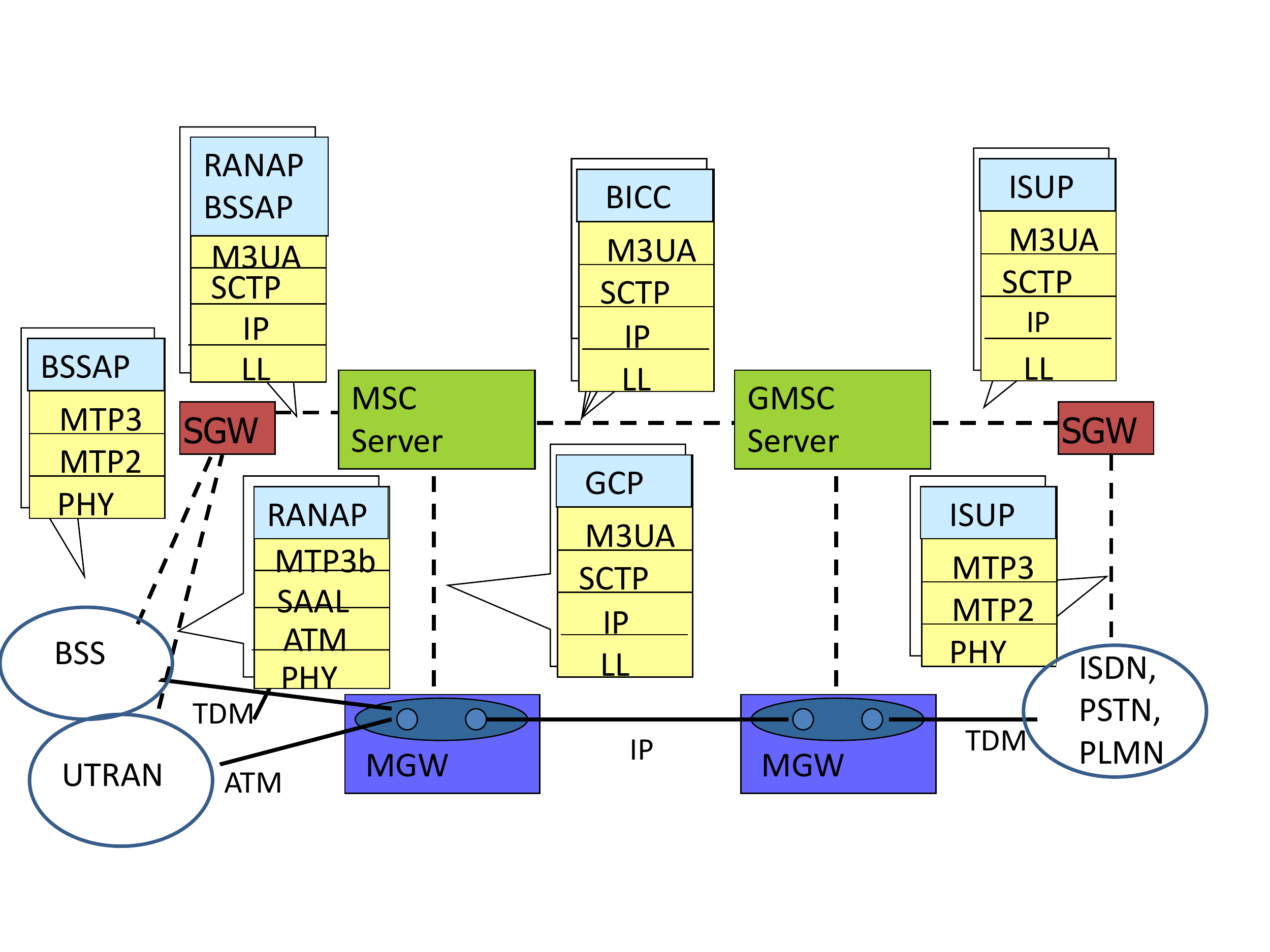}
\caption{Example 3. MSC in signalling network 3G architecture (3GPP standard).} \label{Signalling3gpp}
\end{figure}

In Figure \ref{Signalling3gpp} is presented an example of signalling in mobile core network architecture in 4G of 3GPP. By signalling we mean all the communication needed prior to call establishment and posterior to call release. At the figure all links are represented by protocol stacks that are used to communicate among network nodes. In this particular example we have depiced the case where users from the GSM and UMTS networks (BSS, UTRAN) comunicate over the core network nodes MSC, GMSC by using Media Gateway functions for physical resources, with users located in one of the traditional networks ISDN, PSTN, PLMN. The protocol stacks follows the OSI communication model and functions organised into distict layers are equivalent to functions of each ekvivalent OSI layer. Note that signalling communication is achieved at fourth layer of OSI model.   

\subsection{Hierarchy}
\label{subsect:hierarchy}
Another concept to manage system complexity is to further restrict possible communication interactions by organizing system into tree-like hierarchy.  Communication is allowed only between the modules of the same layer or with a modules of upper or lower layer. Thus, communication possibilities is significantly reduced. Furthermore, in hierarchical systems we may differentiate among communication types but also among roles the end point modules take in communication. For example, we may distinguish between peer--to--peer and client--server communication. In peer--to--peer communication both modules in interaction are equal and both may initiate communication and exchange information. During the whole module lifecycle, both modules are aware of each other. On the other hand side, in client--server type of communication the client side is always requesting some service from the server side while the server side is unaware of possible clients until the service request is  received. The communication may be requested only from the client side and server is serving numerous client requests. These differences in communication types have also reflecting on benefits of system separation on distinct system layers. System layering with hierarchy allows specialisation of layers and their functions and involve hierarchy of functions within the system. 

\subsection{Exercises}

\textit{\textbf{Exercises for students to asses learning's from the chapter just read }
\begin{enumerate}
    \item What are the four main design principles?
    \item Define the term abstraction as design concept for structuring complex systems and provide an example. Discuss benefits.
\item Define the term layering as design concept for structuring complex systems and provide an example. Discuss benefits.
\item Define the term hierarchy as design concept for structuring complex systems and provide an example. Discuss benefits.
\end{enumerate}}
\textit{\textbf{Exercises for students that requires students to reflect on applying design principles in practice}
\begin{enumerate}
    \item Depict structure of the complex system that you imagined as part of excersise at the end of the first section. Use system design principles.
        \item Discuss each design principle you used to structure your system. 
            \item Imagine how your system will evolve in the future. 
            \item Discuss challenges that software organisation may face further evolving such system.
            \item Can you think about adding more abstractions into previously depicted system structure. Discuss benefits.
\end{enumerate}}

\section{Technologies that promote system design principles}
\label{sect:Instruments}

New era of technology is devoted to Information and Communication Technology (ICT). Numerous efforts have been invested into research and technology innovations aiming to provide efficient and effective ways to share information among distant parties that may communicate to each other in timely manner and that these distant parties are not mandatory humans. New technological advances within telecommunication industry and computing industry are driven with this same aim. From telecommunication perspective we are witnesses of revolution. Complete core network infrastructure has been redesigned to cope with these new challenges to carry massive traffic, with very diverse information content, for fixed and mobile users, across various geographic and application domains.

Recent technology advances have been actively promoting  above mentioned design principles. Here we will introduce basic technologies we will use in following section dealing with management functions of virtual network resources. 
The key new technological trends that have seriously impacted the way systems are developed are client--server architecture, system virtualisation technologies and Service Orientation. Understanding these three technologies is prerequisite for getting introduced with new telecommunication era involving Network Virtualisation Functions that is the main subject of this lecture.

\subsection{Restricting communication}
\label{subsection:clientserver}

When an system has modular architecture the it just means that system may be decomposed into number of functional modules. However, amount of communication possibilities is still unlimited and it exponentially grows as system is getting more complex. Propagation of faults in such system is unpredictable and left without any control. Another approach to introduce more control into system operation is to introduce hierarchy into communication processes and more restrictions into communicating rules. This means that design rules are not only documented as guiding principles for designing interfaces but it means also to restrict possible actions over the interface with use of specialised technologies that will enforce interface designers and programmers to use rules and minimise fault propagation within system. Furthermore, change in communication processes should be clearly defined in advance, controlled and independently implemented among affected parties. There are two possible communication types: through protocols and through interface. Protocols are used for example in telecommunication network among peer users that communicate and is represented by a set of defined rules related to information exchanges among communicating parties. Well defined protocols have defined set of messages, set of data exchanged by messages, state behaviour or allowed message sequences, a compatibility mechanism, and timer mechanism. These are the main elements of all standardised protocols.

On the other side, an interface represents a shared boundary (media) for information exchange. Client--server architecture is using interface. Introducing client--server architecture in communication among system modules means to organize modules within the system as clients and services. The communication among the modules is thus restricted. System behaviour is represented as series of events that can be easily tracked to identify root causes for improper system behaviour. Firstly, the clients can ask services for a specific service only using messages. There are no direct procedure calls between clients so implementation errors may be propagated through the system. Errors may propagate exclusively through messages. Moreover, malicious processes may not affect code in systems modules directly. These processes may be introduced only through the messages. Each module in communication is then responsible to verify correctness of the messages. Furthermore, if client is not satisfied with a service it may ask for a service on other place, until it uses standard interface. Functions are getting standardised through the use of standard interface. Open standards are key to promote further standardisation of system functions. The best design principle is to design an interface with assumption that client and server reside on different hardware resources within different physical machines. 

With well defined communication interface, each module can be designed and developed separately, without knowing implementation details of each other. Each module is implemented as it is intended to run on its own physical hardware machine. This programming approach introduces overhead and is weakly reusing benefits of coexistence of functions on the same hardware resources. 

In client--server communication clients requests a service from the server. This communication is stateless, there are no common global states among client and server and no shared memory data structures (i.e. like stack among them) thus making client and server state machines independent. Furthermore, client does not have to trust to the server and all data and messages received can be locally verified by client. Also, if there is no response in reasonable timestamp then client may decide to ask for another service or take proper recovery procedure. This way clients are enforced to be implemented as self--protective, i.e. in case of unexpected behaviour from message oriented interface it can implement regular recovery procedures (i.e. release all related processes before termination of active process or selection of another service in order to successfully complete its function). Furthermore, by using public interfaces the competition among programmers is encouraged so the best implementation may wins. 

\subsection{Service orientation}
\label{subsection:SOA}

Service orientation is a concept where system is made of services implemented as Web services which may be communicated via open standards. These services are dynamically discovered and used on demand, \cite{vliet:softw-eng}. 

\begin{figure}
\centering
\includegraphics[scale=0.46]{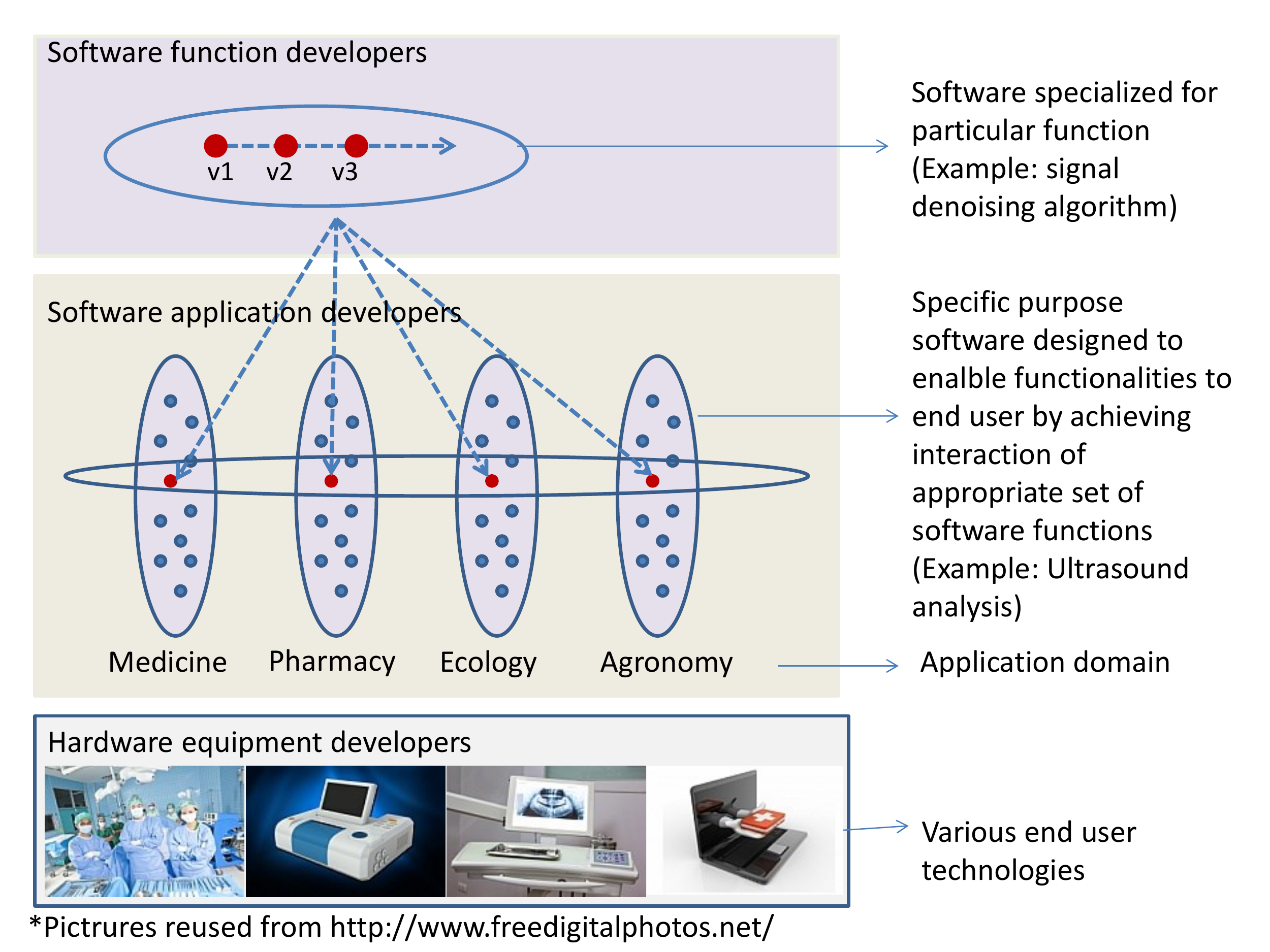}
\caption{Complex systems composed of services.} \label{fig:Complexsystservices}
\end{figure}

There are already many software solutions available in various application domains (medicine, pharmacy, ecology, agronomy, and many others). However, in most of the cases these software is traditionally developed, in vertical and monolithic fashion. In Figure \ref{fig:Complexsystservices} vertical ellipses represent software applications developed for each application domain in traditional monolithic manner and each is tailored to the specifics of the application. These specifics are governed by the specific equipment used in these application domains and are described by industry standards that are often inaccessible to the general public or protected under the intellectual property of individual manufacturers of that equipment. Variety of specialist knowledge were integrated into this single software application and in Figure is represented with a circles. The specialist knowledge (in the form of software) remained locked in the vertical branches of the application domain. However, this knowledge / software may be reused in other domains as well (e.g. signal denoising). But, due to the rigorous monolithic approach to development and the lack of modularity in the software, it is almost impossible to reuse this software/knowledge elsewhere. Furthermore, such monolithic technology limits functional decomposition so it may be difficult to decompose system into separate software functions, but even when possible, it would be difficult to integrate these separate software functions into software applications from other application domains due to the diversity of software technologies used, industry standards, programming languages and  platforms on which they are built. Such, traditional software production imply that the user purchases the software, installs it on his computer, and keeps it working by installing a patch for any errors or being forced to purchase a new version of the software. Such an approach implies ownership of the software, taking care of new versions and evolution, making it much more difficult to maintain. Finally, the biggest concern here is propagation of faults throughout of the system. When failure is occurred, fault is hard to locate, and whole system has to be replaced when introducing changes.

The modular architecture and reusability of already written and tested software has improved significantly in the object-oriented development paradigm. For example, companies in one of the largest domains of the software industry engaged in the development of entertainment applications - software games, achieve significant savings through an object-oriented approach. One character from one game is represented by an object. The characteristics of that character are described by the attributes of that object while the behaviors are described in its functions. This character can be transformed in a number of other games as well. This saves money because all software (code and related documentation) do not have to be developed from scratch and more importantly they have already been tested. However, the interfaces are not standardized and it is difficult for these objects to be reused on other platforms and integrated with other programming languages.

A step further, was the \textbf{component development} approach that made the greatest contribution to interface standardization. The idea behind component development was to develop software components as separate products that implement certain functions that can be offered on the store shelf. By stacking such components, more complex software products can be built. However, the fundamental problem of closed industrial standards was still unsolved and made it very difficult to mix such finished components on heterogeneous technologies. Furthermore, although the components where easy to change and system was modular still there were big concerns about maintaining and upgrading such systems.

\textbf{Service-oriented computing} is a paradigm for the development of complex software applications, which seeks to move computing from the industrial, closed world software production to the open world by using software services independent of proprietary technologies. It is an architectural style of building software applications that promotes loose coupling between components so that you can reuse them and work within a distributed systems architecture. 
Special emphasis is given to:
\begin{itemize}
    \item the transition from the world of industry closed standards to the world of open standards,
    \item The shift from product manufacturing to the use of software services.
\end{itemize}

In the traditional software development paradigm, all software is stored on the personal computer from which it is executed. Or, during the execution of a software application, individual packages are additionally stored and run for execution within a software application already installed. In the new service--oriented paradigm, pieces of software are stored on a distributed network and are called dynamically, on user (client) demand, to perform a specific task. When calling such services, the exact first and last name, number or address of that service is not used. The service is called just based on the description of the client needs. The network offers the service that most closely corresponds to that description at a given moment. This kind of software delivery is reminiscent of customer service, which is why we call these software functions \textbf{services}. Each simple software service is conceived as a software implementation of a certain function that is packaged in such a way that it has a formal and documented interface that is autonomous and loosely--coupled, i.e does not depend on the way other software services and their software implementations are executed. Any simple software service can be located and retrieved, based on open communication norms and mechanisms that are already implemented in the telecommunications network.

The basic difference in this new development paradigm (SOA) is that we do not have all the software on our computer before runtime. Parts of the software are invoked during runtime, and the network makes sure that the user is best served at all times. Maintaining and taking care of new software versions is no longer the subject of headaches for application users. This is what is presented in top of Figure \ref{fig:Complexsystservices}. Since customer is served on a dynamic basis, the network is able to offer the service that best address user needs on the network. Similarly, service descriptions may also contain other information on service quality needs. Software billing is also becoming dynamic, with the use of online services.
Note that the basic prerequisite here is the network that serves the user and the interaction between the user and the network.

\begin{figure}
\centering
\includegraphics[scale=0.33]{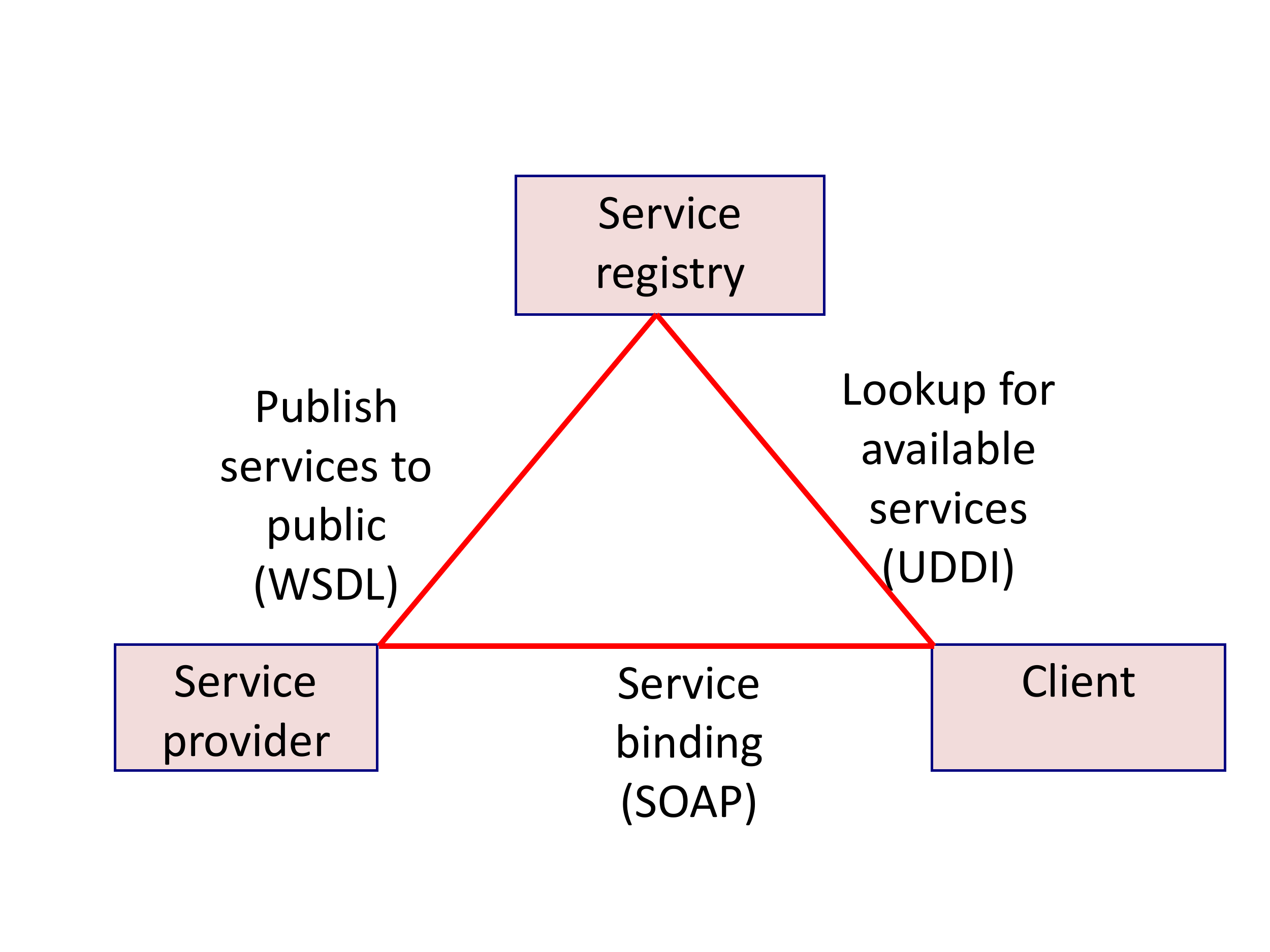}
\caption{Service Oriented Architecture (SOA).} \label{fig:SOA}
\end{figure}

Much of the work in this new paradigm of software development has been transferred to the network, such as storage, retrieval, integration and integration of software in real time -- as needed by software users. This is why the network needed to devise a completely new concept, jointly proposed by OASIS (Organization for Advancement of Structured Information Standards) and the W3C Group (World Wide Web Consortium). This new concept is defined using the well--known Service Oriented Architecture (SOA) presented in Figure \ref{fig:SOA}.
This architecture is based on publicly available and open Internet protocol standards, so it is increasingly cost--effective and simpler to implement. The basis of this paradigm is the Web or Web service, and open standards such as XML language (eXtension Markup Languague) and SOAP (Simple Object Access Protocol), a simple object access protocol used to exchange messages between services. The WSDL (Web Services Description Language) is defined to describe the services, which actually defines the interfaces that access the services and the standard that prescribes the compilation and organization of the Universal Description, Discovery and Integration (UDDI) registry. For the purpose of building software applications based on Web service layouts, the formal language WS - BPEL (Web Service -- Business Process Execution Language) is defined. Some SOA product has been built by many industrial well accepted frameworks but also as part of some virtualisation environments e.g. OpenStack.

There are numerous ongoing research efforts on the analysis of services offered in the network and the evaluation of quality, reliability and security of large and distributed systems based on service-oriented computing. Specifically, in service-oriented computing, it is very important to know how much the quality, reliability and security of the service provided to the user will be, or how a particular Web service will affect the complex software system as a whole. Research is focused on new mechanisms and autonomous monitoring methods and algorithms that will be used for network smart management with the aim of achieving reliable, secure and high-quality service systems. The ultimate goal is full automation, which means that in the future, when matching software systems to the services available, the network will be able to determine for itself which version of a particular service is most appropriate and evaluate the properties of the overall system.

In mobile network evolution, introduction of IMS system into core network has been implemented by using client--server architecture and IMS service offerings as Web services over the REST style API \cite{book:IMS} aiming to link IMS to the Web world and secure IMS services offered over the Internet thus accessible to billions of users. However. Since SOA architecture was under long process of standardisation REST style has been firstly chosen as simpler to implement and easier to use because of numerous available technologies. Further network evolution is integrating SOA for all network functions \cite{osseiran-monserrat-marsch:5G-mob-wireless-comm-tech}.  

\subsection{Virtualisation technologies}
\label{subsection:virtualisation}

Let us firstly define what we mean by saying virtualisation. We can say that we are introducing a new abstraction layer. This actually means separation of upper layer functions from the lower layer functions through new virtualisation layer. Virtualisation layer is mapping the logic from upper layer to lower layer. As already mentioned in hierarchical layering there are strict rules on implementing communication interface among layers. Only neighbouring layers may communicate one to another and only upper layer may request services from lower layer. Then, with such separation these two layers become independent. You can change, duplicate, remove or whatever you want any of these two layers without having to impact the other part. Client--server type of communication is used on interface and introduces additional restrictions that limits error propagation among the layers.

\begin{figure}
\centering
\includegraphics[scale=0.45]{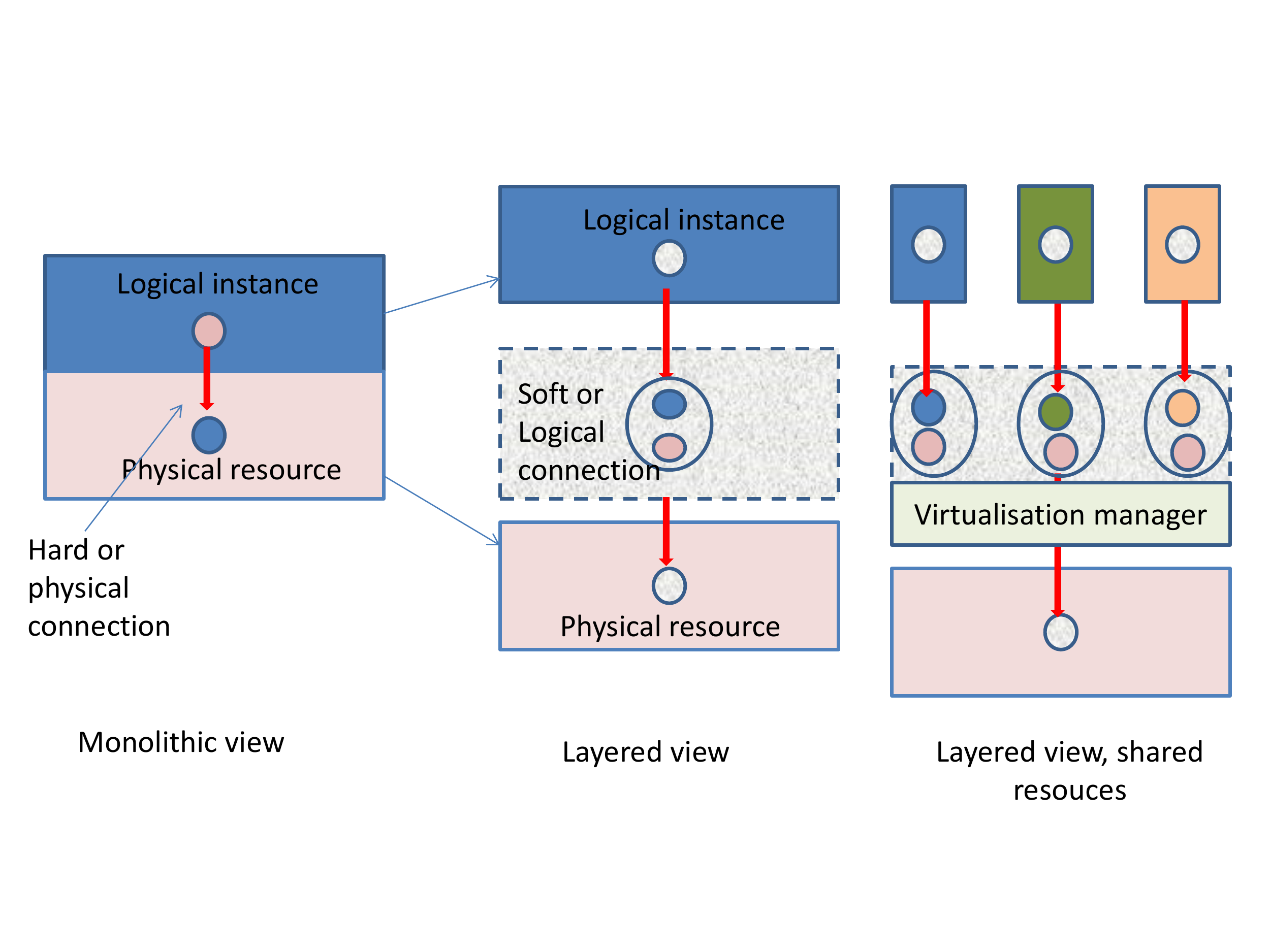}
\caption{Virtualisation concept of managing logical connections.} \label{fig:Virtualisation}
\end{figure}

In the figure \ref{fig:Virtualisation} we present logical decomposition of physical instances in software execution. In classical case, an logical instance is representing a physical resource instance. This connection is fixed and there is no sharing of physical resource instance among various logical instances. When there are multiple calls are running using such an architecture then each logical instance is connected to its own physical instance. Such architecture is inefficient because when the call is inactive the physical resource is left unused. Thus, virtualisation is concept usually applied when we want to gain capacity efficiency and maximal reuse of resources.  

\begin{figure}
\centering
\includegraphics[scale=0.45]{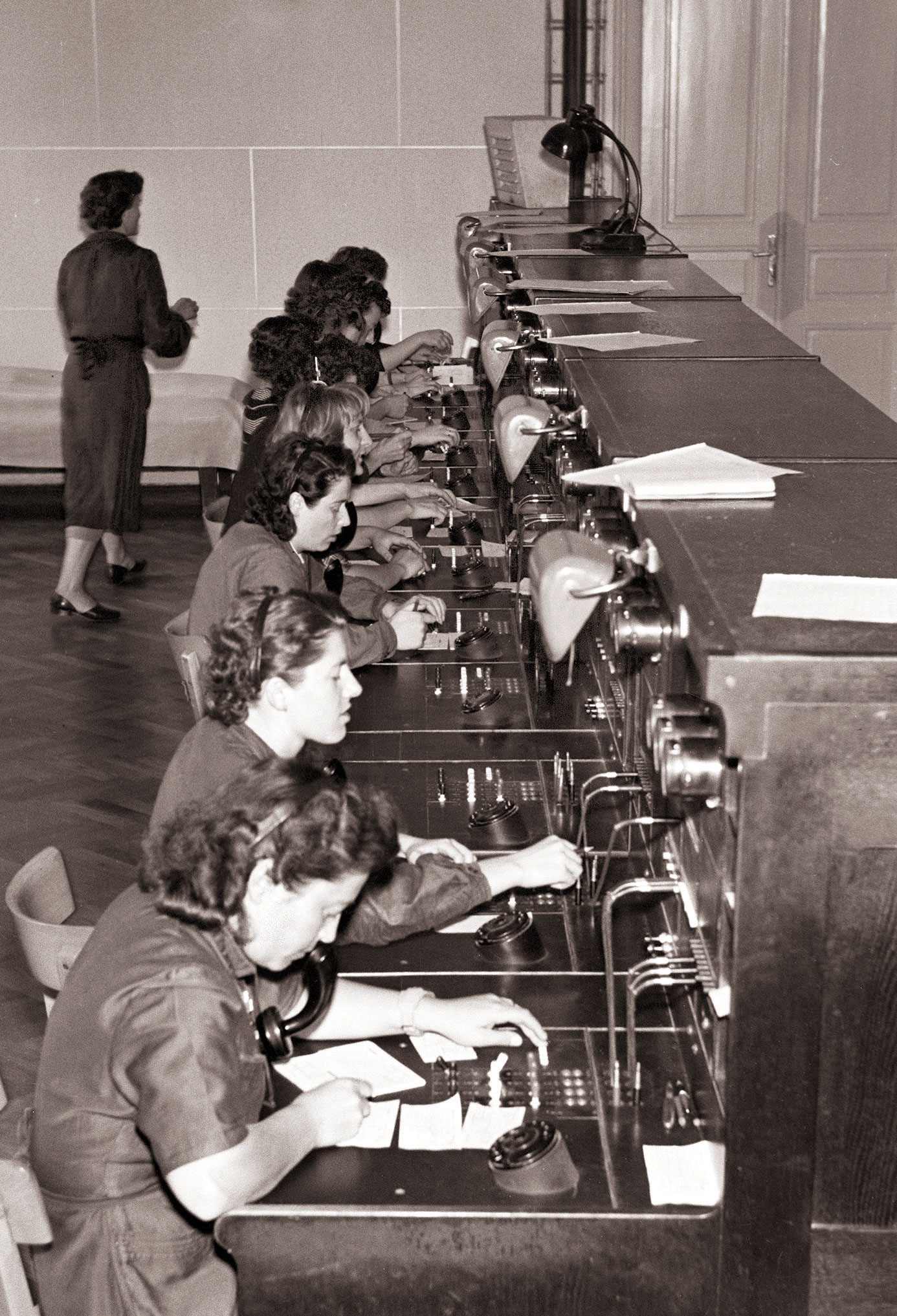}
\caption{First telephone exchange in Maribor, Slovenia 1957.} \label{fig:Firstswitch}
\end{figure}

In telecommunications example we may observe evolution of switching function for telephone calls within an telecommunication exchange. The first switches where physical boards with number of connectors, where each connection is representing one physical wire to one telephone end user and human worker that is manually interconnecting two physical connectors on the switching board. When an telephone user want to subscribe for telecommunication service it gets a physical connector on the switching board in telecommunication exchange. At call establishment phase the caller establish communication with human working on the switching board and ask for connection to callee (a person who is called by caller). Then the human worker at the switch board make a physical link between caller and callee connector at physical board. The physical link is established among caller and callee telephones and they may start conversation on that physical resource. In the Figure \ref{fig:Firstswitch} is presented human workers on switching board in telephone exchange Maribor, Slovenia (1957).

Next step in evolution of switching board was idea of multiplexing number of telephone users over one physical link using time sharing approach. This concept is called time division multiplexing (TDM). This idea involved definition of virtual circuit that is represented by time slot each telephone user is assigned at subscription to telephone service and is called a communication channel. In this phase, each telephone user gets its own virtual channel that is represented as logical instance at logical layer and set of users share one physical wire. This approach introduces sharing of physical resources and there is a need for development of reliable management functions. In this case, still the management of call control processes was completely in control of logical layer. So, error were easily propagating among multiple call processes. 

Introduction of additional virtualisation layer separates management of call control functions from physical resource control functions. All functions related to regular usage of physical resources are left to resource layer management and are separated from the functions related to regular use of logical processes. This separation of call control from resource control logic involves definition of strict rules by using client--server communication principles. Thus the call control and resource control processes are enforced to be independently managed. Common name for new virtualisation layer that is responsible for managing virtual resources is hypervisor. There exists different implementations of hypervisors e.g. in container technology and virtual machine technology. The main difference lies in level of virtualisation involved. Hardware and software resources may be virtualized at various functional layers i.e. hardware level, operating system lavel, library support level and application level \cite{hwang-fox-dongarra:dist-cloud-comp}. While virtual machines are virtualizing resource approach where each virtual instance has its own operating system and fault tolerance and security mechanisms are fully under control of each virtual instance. On the other hand side, containers implements a soft version of virtualisation and provide isolation of containers with underlaying common operating system for all containers running within the same hypervisor. The balance in level of virtualisation has to be find for each application. Virtual machine approach is much more expensive in terms of resource usage but it provides less secure resource control mechanism. For some applications container technology may be sufficient and it is dependent on security requirement for particular application.     

Similarly to case of virtualisation of communication wire where multiple virtual channels are multiplexed in the same communication wire the same virtualisation concept is applied to virtualisation of computer architecture where multiple virtual machines are multiplexed in the same physical hardware machine.

\section{Redesigning the complex system structure}
\label{sect:experience}

In our example Ericsson based MSC node the above mentioned designed principles where introduced with help of key technologies, client--server architecture, service orientation and virtualisation. There are published material explaining in detail modularity concepts introduced and its internal software architecture \cite{AXE810}. These concepts were reused from networking example where each function within the network architecture is defined through set of services node offers, and may be implemented on separate hardware node and by different vendors. So, the communication interfaces and protocols are well specified for their interaction in achieving the network functionalities. Their interaction is governed by these communication protocols and left without any knowledge of their internal structure. Such network architecture enable definition of autonomous service functions that communicate with well-defined interfaces. 

Firstly, virtualisation is introduced to enable separation of application layer functionalities from the resource layer functionalities. Application modules are defined to pack application layer functionality that can be sell to the customers and thus increase value of MSC product. On the other side, there is application platform layer where the non-application specific (or application independent) physical or logical functions, resource specific and resource management functions are located. One example is switching function that is implemented as service within application platform. Services are grouped into modules. The platform is responsible to coordinate common resources. Design principles are introduced specifying that all modules within the system provide its service to other modules or external users over an interface or protocol. Thus, a standard set of communication rules is introduced. The interface among application modules and application platform layer is called application platform service interface (APSI) and contains set of service specifications that are provided over that interface. It is an client--server interface and is independent of service users, does not depend on user implementation and configuration specifics. These service specifications are describing services and the ways how to approach specific service. 

Furthermore, the product structure is documented for the purpose of its easier management. The products from the product structure are somehow categorized and hierarchically layered. In the case of MSC node the product hierarchy is following OSI layer hierarchy as is presented in Figure \ref{subsect:layering}. This structure is represented through product numbering scheme. The benefits of this kind of product management are numerous and are related to collaborative product management. I.e. the strict numbering scheme also implies relationships among modules that are somehow interrelated in specific functionality and product control level and implies any special service agreements. Interfaces also become a part of product portfolio. A set of standards defined that are related to procedures and recovery actions needed while interfaces and modules are changes. 

Such architecture enabled easier product marketing and production of variety of MSC node configurations reusing the same set of software modules. Strict definition of interfaces, restrict fault propagation. Furthermore, independent changes can happen at application layer and application platform layer.

\subsection{Exercises}

\textit{\textbf{Exercises for students to asses learning's from the chapter just read }
\begin{enumerate}
    \item What are the benefits of introducing new abstractions in the system?
    \item When and how we identify need for system restructuring.
\end{enumerate}}
\textit{\textbf{Exercises for students that requires students to reflect on case study}
\begin{enumerate}
    \item What do you think may be limitations and obstacles when introducing new abstractions. See, for example, \cite{AXE810}.  
\end{enumerate}}

\section{Network evolution and further role of functional programming}
\label{sect:discussion}

The main goal during network evolution is to enable different technologies, vendors to access network infrastructure and to provide their interconnection effectively and with minimal use of resources. Network provide an shared resource for all interconnected parties. New generation of networks introduce two concepts: Software Defined Network and Network Function Virtualisation. With these two new concepts the network is opening its resources to be used and configured by its end users and on demand fashion. Furthermore, network introduce self--organisation and autonomic network management functions \cite{agoulmine:autonom-network-manag}. Along this new concepts, existing complex systems have to re--engineer their internal structure so it can provide as much as possible its functions in as a service fashion. Use of open standards is promoted by network infrastructure. New business models will revolutionarize future telecom business. 

In the second part of this lecture we introduce these new technologies, provide reflections on system design principles. We discuss autonomic (networking) design principles on management functions for network operation. Furthermore, we define network design principles that will drive future innovation within the network. Finally, in this new generation of network a shift will be made from system programming to network programming. Here as well, the functional programming approach would be enforced.

\end{document}